\documentstyle[floats,aps,prl,epsf,twocolumn]{revtex}

\begin{document}
\draft

\wideabs{

\title{Charge Order driven Spin-Peierls
Transition in $\alpha'$-Na$_{x}$V$_2$O$_{5}$}

\author{Y.~Fagot-Revurat$^{1}$, M.~Mehring$^{1}$, R.~K.~ Kremer$^{2}$}
\address{$^1$ 2. Physikalisches Institut, Universit\"at Stuttgart, 70550
         Stuttgart, Germany}
\address{$^2$ Max-Planck-Institut f\"ur Festk\"orperforschung, Heisenbergstrasse
        1, D-70569 Stuttgart, Germany}

\date{\today}

\maketitle

\begin{abstract}
We conclude from $^{23}$Na and $^{51}$V NMR measurements in
$\alpha'$-Na$_{x}$V$_2$O$_{5}$ ($x$=0.996) a charge ordering transition
starting at $T\leq 37~K$ and preceding the lattice distortion and the
formation of
a spin gap $\Delta$=106 K at $T_c$=34.7~K. Above $T_c$, only a single Na site
is observed in agreement with the $Pmmn$ space group of this first 1/4-filled
ladder system. Below $T_c$=34.7~K, this line evolves into eight distinct
$^{23}$Na quadrupolar split lines, which evidences a lattice distortion with,
at least, a doubling of the unit cell in the $(a,b)$ plane. A model for this
unique transition implying both charge density wave and spin-Peierls order is
discussed.

\end{abstract}
\pacs{PACS numbers: 76.60.-k, 74.25.Ha, 74.62.Dh}

}

\narrowtext

A large number of vanadium-oxides have recently been recognized as
low-dimensionnal quantum spin systems with outstanding magnetic properties. In
several cases, and at odd with the experience in copper-oxides, defining the
parameters of the magnetic Hamiltonian has turned out to be more complicated,
and more intriguing than originally anticipated, which has led to a number of
surprises \cite{V-oxides}. The most striking example is
$\alpha'$-NaV$_{2}$O$_{5}$. The $P2_{1}mn$ space group first proposed for its
room-temperature crystal structure in 1975 \cite{Carpy75} allowed the
discrimination between two distinct V sites in the unit cell, rationalizing the
1D magnetic properties: Namely, chains of V$^{4+}$ (3d$^{1}$) ions,
antiferromagnetically (AF) coupled along the b-axis with an exchange constant
$J$=430-560 K, are decoupled by spinless V$^{5+}$(3d$^{0}$) chains
\cite{Isobe96,Fuji97}. Three years ago, the discovery of a doubling of the
lattice periodicity below $T_c$=35 K \cite{Isobe96} associated with the opening
of a gap $\Delta$=9.8 meV in the magnetic excitations \cite{Fuji97}, was first
analyzed as the signature of a spin-Peierls (SP) transition \cite{Bray83},
which would be the second case for inorganic solids, after CuGeO$_{3}$
\cite{Hase93}.

Nevertheless, several arguments can be raised against a simple SP-transition in
this system : the strong BCS ratio 2$\Delta_0/k_B T_c$=6 \cite{Fuji97}
(compared to 3.5 in other SP systems), the weak field dependence of $T_{c}$ (5
times too small) \cite{Schnelle99} and the unusual distortion wave vector
$k_c$=(1/2,1/2,1/4) \cite{Fuji97}. In fact, recent X-ray investigations have
revealed a centrosymmetric Pmmn space group (SG Pmmn) for the crystal structure
with only one V site per unit
cell \cite{Smolinski98,Schnering98,Meetsma98}. This has led to a redefinition
of $\alpha'$-NaV$_{2}$O$_{5}$ as the first experimental realization of a
1/4-filled ladder system made of weakly coupled V$^{4.5+}$ chains, where a spin
1/2 is carried by the V-O-V orbitals on the rungs of the ladder (one
$d$-electron per two equivalent V ions). This configuration leads to an
effective S=1/2 1D Heisemberg AF spin Hamiltonian due to the strong AF coupling
between the rungs \cite{Smolinski98,Horsch98,Seo98}. The most striking aspect
of this new type of quantum spin chain is the strong interplay expected between
lattice, spin and charge degrees of freedom, which should lead to peculiar
instabilities \cite{Thalmeier99,Riera98}.
Below $T_c$, recent NMR results have proved the existence of two different V
sites, attributed to V$^{4+}$ and V$^{5+}$\cite{Ohama99}. This has motivated
recent theoretical works taking into account a pure charge ordering (CO)
instability \cite{Mostovoy98,Seo98} or CO coupled to SP-distortion
\cite{Thalmeier99,Riera98}. However, experimental proof in the vicinity
of the transition is urgently required.

In this Letter, we present a $^{23}$Na ($I$=3/2) NMR study in a
well-characterized single crystal of $\alpha'$-Na$_{x=0.996}$V$_2$O$_{5}$
\cite{note1}. The strong quadrupolar interaction between $^{23}$Na nuclei and
surrounding ions allows to monitor the structural transition occurring at $T_c$
=34.7 K. Furthermore, by combining $^{23}$Na and $^{51}$V NMR, we infer a {\it
charge ordering} of V$^{4.5+}$ into V$^{4.5-\delta}$ and V$^{4.5+\delta}$
($\delta\ll$0.5) for T$<37$~K, {\it i.e. preceding the lattice
distortion}. This unconventional double transition implies the presence of both
CO and SP ordering.

\begin{figure}[!t]
\centerline{\epsfxsize=80mm \epsfbox{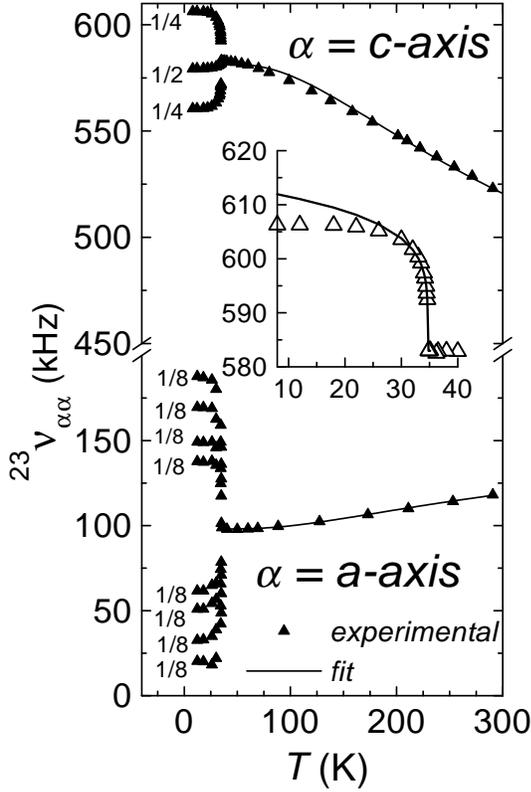}}
\caption{$T$-dependence of $^{23}\nu_{\alpha\alpha}$ measured
for $H_{0}//\alpha$=$a$ and $c$-axis (the continuous line is a fit explained
in the text), Inset : detail of the 2nd order 
transition occuring at $T_{c}$=34.7 K for $H_{0}//$$c$ 
(the line is a fit to a critical law with $\beta$=0.19).}
\label{fig1}
\end{figure}

The three diagonal components of the quadrupolar tensor are given by
$\nu_{\alpha\alpha}$=$eQV_{\alpha\alpha}/2h$ \cite{Slichter90} with principal
axes $\alpha$=Z,Y,X, $e$ the elementary charge, $Q$ the electric quadrupole
moment, $h$ the Planck constant and $V_{\alpha\alpha}$ the principal values of
the electric field gradient (EFG) tensor at the Na site. They can be readily
extracted from the experimental data, free of the magnetic contribution, by
measuring
[$\nu_{H_{0}||\alpha}$(-1/2,-3/2) -
$\nu_{H_{0}||\alpha}$(3/2,1/2))]/2=$\nu_{\alpha\alpha}$
where $\nu_{H_{0}||\alpha}(\pm 1/2,\pm 3/2)$ are the satellite lines for
H$_{0}||\alpha$=X,Y,Z. Since the EFG tensor is traceless the two parameters,
$\nu_{ZZ}$ and $\eta$=($\nu_{XX}$-$\nu_{YY}$)/$\nu_{ZZ}$) (asymmetry parameter)
represent the EFG tensor fully. The $^{23}$Na NMR measurements were carried out
by Fourier transforming the spin-echo signal recorded in a magnetic field
H$_{0}$=4.26 Tesla. According to the local symmetry of the Na site in the SG
Pmmn
, the a, b and c-axis are unambigously identified as the principal axes
and we deduced the parameters $\nu_{ZZ=bb}$=-641 kHz (-678 kHz), $\nu_{XX=cc}$=
522 kHz (581 kHz), $\nu_{YY=aa}$=118 kHz (99 kHz) and $\eta$=0.630 (0.711) for
T=290 K (resp. $T$=50 K) (see fig. 1) in agreement with earlier powder data at
50 K \cite{Ohama97a}. We note that the principal axes (XYZ) correspond to the
crystal axes as (X=c,Y=a,Z=b). The EFG tensor can be expressed as
V$_{total}$=[1-$\gamma_{\infty}$]V$_{lattice}$+[1-R$_{Q}$]V$_{ion}$
with V$_{ion}$ corresponding to the ion under investigation, and where
$\gamma_{\infty}$ and R$_{Q}$ are empiric antishielding
factors. V$_{lattice}$ defines the pure lattice contribution and is given by
\begin{equation}
V_{lattice}=V_{\alpha\beta}=\sum_{j}q_{j}[3X_{\alpha}^{j}X_{\beta}^j-\delta_
{\alpha\beta}r_{j}^{2}]/r_{j}^{5}
\end{equation}
where q$_{j}$ is the charge at site j, X$_{\alpha,\beta}^j$ is the projection
of the vector $\vec{r_j}$ from the nucleus under consideration to the charge
$q_j$. Applying a ionic point charge model by using the atomic positions of the
SG Pmmn \cite{Meetsma98} and the average valencies Na$^{1+}$, O$^{2-}$
and V$^{4.5+}$ and V$_{ion}$=0 (in the Na case) and by setting
$\gamma_{\infty}$=0, results in a lattice contribution of $\nu_{bb}$=-492 kHz,
$\nu_{cc}$=+469 kHz and $\nu_{aa}$=+22 kHz, which confirms our labeling of the
axes. The main contribution to $\nu_{\alpha\beta}$ comes from the oxygen
(d$_{Na-O}$=2.6 \AA) and vanadium (d$_{Na-V}$=3.6 \AA) atoms. The agreement is
rather good but deviations are of course expected at short distances due to
covalency effects neglected in the ionic point charge approximation. Applying
the same point charge model to the P2$_{1}$mn structure proposed in
\cite{Carpy75} leads to rotated principal axes in the a-c plane inconsistent
with the experimental data. Thus, our observation of only one Na site at room
temperature supports the SG Pmmn (on the NMR time scale \cite{note2})
in agreement with X-rays, Raman and $^{51}$V NMR data
\cite{Smolinski98,Schnering98,Weiden98,Ohama99}.

\begin{figure}[!t]
\centerline{\epsfxsize=80mm \epsfbox{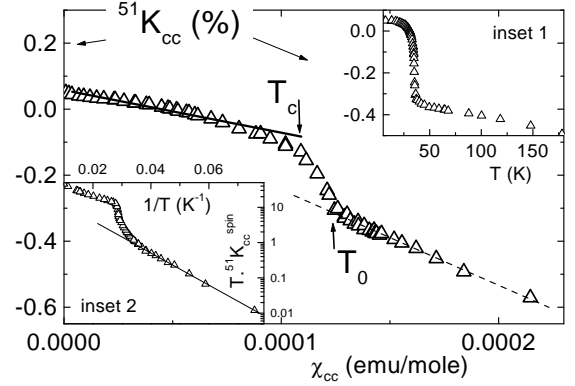}}
\caption{$^{51}K$ as a function of $\chi^{spin}$ with $T$ as an intrinsic
parameter.
Inset 1: $T$-dependence of $^{51}K$ at the V$^{4.5+}$ site
for $T$$>T_{0}$=37 K and at the V$^{4.5+\delta}$ for $T$$<T_{c}$=34.7 K.
Inset 2:
linear variation of $T$$^{51}$K as function of 1/$T$ leading to a gap
$\Delta$=106 K.}
\label{fig2}
\end{figure}

In order to check for modifications of V$_{lattice}$ due to any change
of the charge configuration, we have measured the $T$-dependence of 
$\nu_{aa,bb,cc}$ between 8 and 300 K (fig. 1) . We can divide the results into 
two different
regimes. A high temperature regime for $T\gg T_{c}$=34.7 K (fig. 1), where an
increase of $\nu_{bb}$ and $\nu_{cc}$ (a decrease of $\nu_{aa}$) was observed
when decreasing $T$. This is attributed to the decrease of the amplitude of
lattice vibrations: The best fit (for H$_{0}||$a-axis) to the Bayer formula
\cite{Bayer51}
$\nu_{\alpha\alpha}$(T)=$\nu_{\alpha\alpha}$(T=0)[$\gamma+[\beta/(exp(
\Theta_D/T)-1)$]] with $\gamma$=0.99, $\beta$=0.25, $\nu_{aa}$(T=0)=99.1 kHz
leads to a Debye temperature $\Theta_D$=358$\pm$ 13 K in good
agreement with reference \cite{Schnelle99}.

Below T$_{c}$=34.7~K, a pronounced splitting of the Na satellite lines is
observed for all magnetic field orientations (fig. 1), whereas no drastic
change has been observed on the central line, thus ruling out any changes in
magnetic interactions proving this to be a pure quadrupolar effect. Each
satellite line is split into two sets of four lines of the same intensity (1/8)
corresponding to 8 different quadrupolar Na sites in the low-T phase for
H$_0||$a (fig. 1) and H$_0||$b (not shown here). In contrast only 3 lines are
visible for H$_0||$c with half of the intensity for the central site and 1/4
for each of the two other sites. This is readily explained as a superposition
of different sites for H$_0||$c when rotating the sample
in the a-c plane (not shown here). For T$<$T$_{c}$,
$\nu_{\alpha\alpha}$(T) follows a critical behavior (for H$_0||$a,b or c-axis)
corresponding to a second order transition connected with a continuous change
of the atomic positions of vanadium and/or oxygen atoms as determined by X-rays
\cite{Fuji97}. Such a behavior is demonstrated here for one site for H$_0||$c
in the inset of fig. 1. We used a fit following the relation
$\nu$(T)=$\nu(T_{c}$)+$\delta\nu[1-T/T_{c}]^{\beta}$ resulting in the
parameters $\nu(T=T_{c})$=582.96 kHz, $\delta\nu$=30.5$\pm$0.9
kHz,T$_{c}$=34.66$\pm$0.04 K and a critical exponent $\beta$=0.19$\pm$0.01 in
reasonable agreement with $\beta$=0.25 and 0.20 as measured from the phonon
anomaly \cite{Smirnov98} and by X-rays \cite{Ravy99}. This critical exponent is
far from the expected mean field value ($\beta$=0.5) but also far from the one
obtained in other SP compounds ($\beta$=0.32) {\it indicating clearly that this
transition in $\alpha'$-Na$_{x}$V$_2$O$_{5}$ is of a different type.}
According to the 8 distinct quadrupolar sites observed below T$_{c}$ in the a-b
plane, the unit cell is at least doubled along the a and b-axis in agreement
with the wave vector k$_{c}=(1/2,1/2,1/4)$ proposed in \cite{Fuji97}.
Considering the changes in V$_{lattice}$ ($\delta\nu_{cc}/\nu_{cc}$=4 \%,
$\delta\nu_{bb}/\nu_{bb}$=8 \%, $\delta\nu_{aa}/\nu_{aa}$=85 \%) as derived
from our measurements, the displacements occur
unusually mainly along the a-axis. In addition, we
found that applying only charge disproportionation at the V sites does not
significantly affect the calculated EFG tensor at the Na site.
{\it We conclude that $^{23}$Na NMR probes a pure second order lattice
distortion
at T$_{c}$=34.7 K}.

\begin{figure}[!t]
\centerline{\epsfxsize=80mm \epsfbox{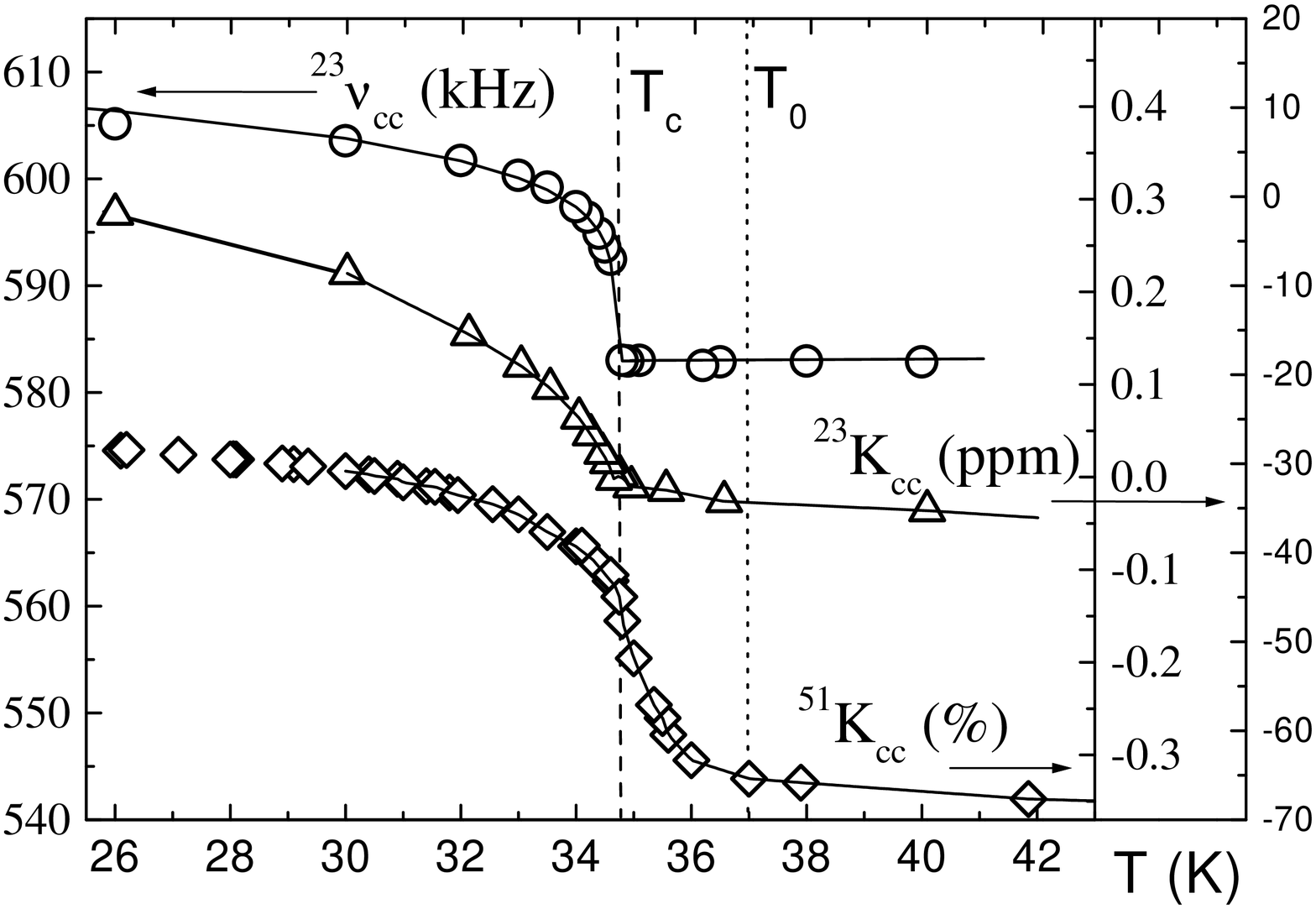}}
\caption{ $T$-dependence of $^{23}\nu_{cc}$, $^{23}$K and
$^{51}$K in the vicinity of the double transition.}
\label{fig3}
\end{figure}

In addition, we have performed $^{51}$V NMR in the same single crystal
for H$_{0}||$c, in order to check on the change of
the electronic state of the vanadium atoms when going through the transition.
Only one V site (one set of 7 lines for a
nuclear spin $I$=7/2) was observed at $T$=100 K, with the parameters
$\nu_{cc=ZZ}$=552 kHz and $\eta$=0.24, in agreement with Ref.
\cite{Ohama99}. The magnetic
hyperfine shift $^{51}$K$_{cc}$ at this vanadium site (fig. 2) was
extracted from the $T$-dependence of $^{51}\nu(-1/2,1/2)$ between 5 and
300 K (inset 1, fig. 2). The shift measured for H$_{0}||\alpha$ can be expressed
as :
\begin{equation}
K_{\alpha\alpha}(T)=A_{\alpha\alpha}
\chi_{\alpha\alpha}(T)/\mu_{B}N_{A}g_{\alpha\alpha}
+K_{\alpha\alpha}^{orb}
\end{equation}
where A$_{\alpha\alpha}$ is the hyperfine coupling in Oe, $\chi_{\alpha\alpha}$
is the magnetic suceptibility in emu/mole, $\mu_{B}$ is the Bohr magneton in
erg/Oe, N$_{A}$ the Avogadro number in mole$^{-1}$, g$_{\alpha\alpha}$ the
corresponding $g$-tensor component and K$_{\alpha\alpha}^{orb}$ is the orbital
contribution. We first attribute the slow decrease of $^{51}$K$_{cc}$ above
T$_{c}$ followed by a rapid drop around T$_{c}$ (inset 1, fig. 2) to the change
of the magnetic suceptibility as is conceivable for a $S$=1/2 Heisenberg AF
chain approaching a singlet ground state through a SP-transition. In the $T$=0
limit, $\chi$ vanishes, which allows us to determine $^{51}K_{cc}^{orb}$=0.063
\%. In order to determine A$_{cc}$ we have plotted $^{51}K_{cc}(T)$ as a
function of $\chi(T)$ with $T$ as an implicit parameter (fig. 2). Usually,
this leads to a unique linear relation in which the slope corresponds to the
hyperfine coupling and the $T$=0 limit is the orbital shift \cite{Fagot97}.
Suprinsingly, we observe here a cross-over in the range 34.7 K$<$T$<$37 K
separating two distinct linear regimes. Two sets of parameters (A$_{cc}$=-78.5
kOe, K$_{cc}^{orb}$=0.029 \% for T$>$T$_{0}$=37 K) and (A$_{cc}$=-41.5 kOe,
K$_{cc}^{orb}$=0.063 \% for T$<$T$_{c}$=34.7 K) are deduced, indicating a
strong modification of the electronic state at the vanadium site in the
vicinity of the transition. In the low-$T$ phase, we have detected two sets of
$^{51}$V lines, but only the one with longer relaxation time and narrower
linewidth is discussed here in detail, close to the transition \cite{rem2}. The
full spectrum has been recorded for this site at $T$=12 K and we deduced
$\nu_{cc=ZZ}$=399 kHz, $\eta$=0.5 \cite{rem2}. Referring to the parameters
obtained previously for T$<$T$_{c}$ (A$_{cc}$=-31 kOe, K$_{cc}^{orb}$=0.063 \%,
$\nu_{cc=ZZ}$=399 kHz, $\eta$=0.5), we are now able to correlate this site with
the one identified as the V$^{5+}$ site in Ref. \cite{Ohama99}. This confirms
charge ordering from only a single V site with valency V$^{4.5+}$ above T$_c$
to two different sites, a nominally V$^{5+}$ (shown here) and 
V$^{4+}$ site (not shown here) below T$_c$(see discussion below). Applying the
relation T$\chi^{spin}$(T)$\propto$ exp(-$\Delta$/T) for a dimerized state
\cite{Bulaevskii69}, subtracting the orbital contribution, we deduce a
spin gap with $\Delta$=106.3 (1.3) K (inset 2, fig. 2) in good agreement with
the values already derived from other
data\cite{Fuji97,Ohama97a,Ohama99,Luther99,Kuroe99}.

Noting that (i) $\nu_{ZZ}$=399 kHz attributed to the nominal
V$^{5+}$ site is larger than the one measured for the pure V$^{5+}$ site in
V$_{2}$O$_{5}$ ($\nu_{ZZ}$=63 kHz) \cite{Ohama97a}, and (ii) we have measured a
finite spin gap at this site. Thus this cannot be a pure V$^{5+}$ state but
rather an hybridised V$^{4.5+\delta}$ site (resp. V$^{4.5-\delta}$ for the
other one), with $\delta$$\ll$0.5 varying slightly with temperature.

Fig. 3 summarizes the $T$-dependence of $^{51}$K$_{cc}$, reflecting the charge
ordering at the V sites, of $^{23}\nu_{cc}$, reflecting the lattice distortion
and of $^{23}$K$_{cc}$, which evidences the opening of the spin gap (scaling
perfectly with $\chi$(T)). Here, we want to point out an important new
observation not noticed before : It follows from fig. 3 that the charge
ordering, evidenced by the rapid drop of $^{51}$K$_{cc}$ in the range $34.7
K<T_{0}<37 K$ {\it precedes} the lattice distortion and the opening of the spin
gap, both occuring at T$_{c}$=34.7 K. The linearity observed between $\chi(T)$
and $^{51}K_{cc}(T)$ (fig. 2), already holds at T$_{c}$=34.7 K, and proves that
this phase transition sets in only when the V$^{4.5+\delta}$ site is already in
a new electronic configuration. In fact, this was proved directly by
recording the low frequency $^{23}$Na satellite line and the
$^{51}$V$^{4.5/4.5+\delta}$ central line {\it simultaneously} (their resonance
frequencies are very near to each other close to T$_{c}$): modifications of the
$^{51}$V NMR spectrum between 37 and 34.7 K are not accompanied by any change
of $^{23}\nu_{cc}$. We can also exclude the possibility of multiple transitions
(as suspected in high resolution thermal expansion measurements
\cite{Koppen98}) by considering the sharp C$_{p}$ anomaly observed in crystals 
of the same batch\cite{Schnelle99} and by recalling that we probe, in the
same sample,
two different physical quantities 
on two different nuclei, both showing clearly distinct behaviors.

At this stage, we are ready to discuss (i) the nature of the transition and (ii)
the corresponding low-$T$ structure. From our results, we can exclude a pure
charge ordering as proposed in \cite{Mostovoy98,Seo98}. This would lead to the
simultaneous occurence of lattice distortion and charge ordering in a single
transition. On the other hand, the CO observed for T$_{0}<37 K$ 
excludes a pure SP ordering in agreement with the analysis of
\cite{Hemberger98}. The presence of CO as a precursor of the transition could
be an explanation for the weak field dependence of T$_{c}$ \cite{Schnelle99}
and the strong BCS ratio \cite{Fuji97}, as proposed in \cite{Thalmeier99}.
Finally, recent numerical results point out the strong interplay between charge
density wave instability and spin-Peierls order as intrinsic to this 1/4-filled
ladder system and this should lead to a transition involving both lattice, spin
and charge degrees of freedom \cite{Riera98}. This is in good agreement with
the double transition experimentally observed here.
The simplest scenario consists in first a charge ordering restoring in part the
magnetic decoupling between adjacent chains (for $34.7 K<T_{0}<37 K$), followed
by a dimerisation at T$_{c}=34.7 K$ with the formation of singlets along the
b-axis. The low-$T$ structure should consist of dimerized chains of
V$^{4.5-\delta}$ partially decoupled by inert chains of V$^{4.5+\delta}$ as
proposed in \cite{Thalmeier99}. This simple ordering should however be
energetically less favorable than a zig-zag arrangement between adjacent
ladders \cite{Riera98,Mostovoy98,Seo98}.

Clearly, a better experimental characterization of the charge order is the key
to understand the magnetism of $\alpha'$-Na$_{x}$V$_2$O$_{5}$. Thus, a
determination of the low-$T$ structure is definetely needed, which must be
compatible with the NMR data, in order to unravel the configuration of the
spin-singlets. The NMR results reported here, with eight inequivalent
quadrupolar $^{23}$Na sites and two different Vanadium valencies, 
and where the lattice distorsions occur predominantly along the a-axis
provide a stringent test for the low-$T$ structure.

We would like to acknowledge X. Bourdon, S. Kr\"amer,
F. Mila and D. Poilblanc for helpful
discussions and M.-H. Julien for his critical reading of the manuscript.
Y.F.-R. was supported by a postdoctoral grant from the Alexander
Von Humboldt Institut (AVH). M.M. would like to acknowledge support by the
Fonds der Chemischen Industrie.

\begin{references}
\bibitem{note1} The transition is supressed for x$<$0.97. Our sample is the one
        having the sharper C$_{p}$ anomaly (no.4) in \cite{Schnelle99}.
\bibitem{note2} The broken-parity electronic ground state observed in
\cite{Damascelli98a}
        could be attributed to slow charge fluctuations compared to the
frequency used as proposed
        in \cite{Nishimoto98}.
\bibitem{rem2} Detailed data will be published in a separate paper.
\end{references}
\end{document}